\documentclass[preprint,showpacs,preprintnumbers,amsmath,amssymb,prb]{revtex4}

\usepackage{epsfig,graphicx}
\usepackage{dcolumn}
\usepackage{bm}
\usepackage[normal]{subfigure}
\usepackage{color}

\usepackage{booktabs}
\usepackage{multirow}
\usepackage{array}
\newcolumntype{+}{>{\global\let\currentrowstyle\relax}}
\newcolumntype{^}{>{\currentrowstyle}}

\newcommand{\me}{\mathrm{e}}

\newcommand{\zerov}{{\bf 0}}
\newcommand{\qv}{{\bf q}}

\newcommand{\Qv}{{\bf Q}}
\newcommand{\kv}{{\bf k}}

\newcommand{\kvp}{\ensuremath{{\bf k^{\prime}}}}

\newcommand{\Kv}{{\bf K}}

\newcommand{\rv}{{\bf r}}

\newcommand{\rvp}{\ensuremath{{\bf r^{\prime}}}}

\newcommand{\rhov}{\ensuremath{\mbox{\boldmath${\rho}$}}}
\newcommand{\wavev}{wave-vector}   %
\newcommand{\wavef}{wave-function}   %
\newcommand{\wavevs}{wave-vectors}   %
\newcommand{\wavefs}{wave-functions}   %
\begin{document}

\title{Single Particle Transport in Two-dimensional Heterojunction Interlayer Tunneling Field Effect Transistor}

\author{Mingda (Oscar) Li$^a$, David Esseni*, Gregory Snider, Debdeep Jena, and Huili Grace Xing$^b$}
\affiliation{%
University of Notre Dame, IN, USA \\
*University of Udine, Italy \\
E-mails: a. mli7@nd.edu; b. hxing@nd.edu
}%

\date{\today}
\begin{abstract}

The single particle tunneling in a vertical stack consisting of
monolayers of two-dimensional semiconductors is studied
theoretically and its application to a novel Two-dimensional
Heterojunction Interlayer Tunneling Field Effect Transistor
(Thin-TFET) is proposed and described. The tunneling current is
calculated by using a formalism based on the Bardeen's transfer
Hamiltonian, and including a semi-classical treatment of scattering
and energy broadening effects. The misalignment between the two 2D
materials is also studied and found to influence the magnitude 
of the tunneling current, but have a modest impact on its gate
voltage dependence. 
Our simulation results suggest that the Thin-TFETs can achieve very
steep subthreshold swing, whose lower limit is ultimately set
by the band tails in the energy gaps of the 2D materials produced by energy broadening. The Thin-TFET is thus
very promising as a low voltage, low energy solid
state electronic switch.

\end{abstract}
\pacs{}
\keywords{Suggested keywords}

\maketitle

\section{Introduction}\label{Sec:introduction}

The electronic integrated circuits are the hardware backbone of
today's information society and the power dissipation has recently
become the greatest challenge, affecting the lifetime of existing portable equipments, the sustainability of large and
growing in number data centers, and the feasibility of energy
autonomous systems for ambience intelligence
\cite{rabaey2002picoradios,amirtharajah1998self}, and of sensor
networks for implanted monitoring and actuation medical devices
\cite{dreslinski2010near}. While the scaling of the supply voltage,
V$_{DD}$, is recognized as the most effective measure to reduce
switching power in digital circuits, the
performance loss and increased device to device variability are a
serious hindrance to the V$_{DD}$ scaling down to 0.5 V or below.

The voltage scalability of VLSI systems may be significantly
improved by resorting to innovations in the transistor technology
and, in this regard, the ITRS has singled out Tunnel filed effect transistors (FETs) as the
most promising transistors to reduce the sub-threshold swing, SS,
below the 60 mV/dec limit of MOSFETs (at room temperature), and thus
to enable a further V$_{DD}$ scaling
\cite{itrs2011international,seabaugh2010low}.
Several device architectures and materials are
being investigated to develop Tunnel FETs offering both an attractive on
current and a small SS, including III-V based transistors possibly
employing staggered or broken bandgap heterojunctions
\cite{zhou2012novel, tomioka2013sub, knoll2013inverters, mohata2011demonstration},
or strain engineering \cite{conzatti2012strain}. Even if encouraging
experimental results have been reported for the on-current in III-V
Tunnel FETs, to achieve a sub 60 mV/dec subthreshold swing is still a
real challenge in these devices, probably due to the detrimental
effects of interface states
\cite{zhou2012novel,pala2013interface,esseni2013interface}.
Therefore, as of today the investigation of new material systems and
innovative device architectures for high performance Tunnel FETs is
a timely research field in both the applied physics and the electron
device community.

In such a contest, two-dimensional (2D) crystals attract increasingly more attention primarily due to their scalability, step-like density of states and absence of broken bonds at interface. They can be stacked to form a
new class of tunneling transistors based on an interlayer tunneling
occurring in the direction normal to the plane of the 2D materials.
In fact tunneling and resonant tunneling devices have been recently
proposed \cite{feenstra2012single}, as well as experimentally
demonstrated for graphene-based transistors
\cite{britnell2013resonant,britnell2012atomically}.
Furthermore, monolayers of group-VIB transition metal
dichalcogenides MX$_2$ (M = Mo, W; X = S, Se, Te) have recently
attracted remarkable attention for their electronic and optical
properties \cite{radisavljevic2011single, wang2012electronics}. Monolayers of transition-metal dichalcogenides (TMDs)
have a bandgap varying from almost zero to ~2 eV with a sub-nanometer thickness such that these materials can be considered approximately
as two-dimensional crystals \cite{gong2013band}. The sub-nanometer thickness of TMDs can
provide excellent electrostatic control in a vertically stacked
heterojunction. Furthermore, the 2D nature of such materials make
them essentially immune to the energy bandgap increase produced by the
vertical quantization when conventional 3D semiconductors are
thinned to a nanoscale thickness, and thus immune to the
corresponding degradation of the tunneling current density
\cite{jena2013tunneling}.
Moreover, the lack of dangling bonds at the surface of TMDs may
allow for the fabrication of material stacks with low densities of
interface defects \cite{jena2013tunneling}, which is another
potential advantage of TMDs materials for Tunnel FETs applications.

In this paper we propose a two-dimensional heterojunction interlayer
tunneling field effect transistor (Thin-TFET) based on 2D semiconductors and develop a transport model based on the transfer-Hamiltonian
method to describe the current voltage characteristics and discuss,
in particular, the subthreshold swing. In Section \ref{Sec:Model} we
first present the device concept and illustrate examples of the
vertical electrostatic control,
then we develop a formalism to calculate the tunneling current.
Upon realizing that the subthreshold swing of the Thin-TFET is
ultimately determined by the energy broadening, in
Sec.\ref{Sec:broadening} we show how this important physical
factor has been included in our calculations. In
Sec.\ref{Sec:misalignment} we address the effect of a possible
misalignment between the two 2D semiconductor layers, while in
Sec.\ref{Sec:AnalyticResults} we derive some approximated,
analytical expressions for the tunneling current density, which are
useful to gain insight in the transistor operation and to guide
the device design. In Sec.\ref{Sec:Results} we present the results of
numerically calculated current voltage characteristics for the
Thin-TFET, and finally in Sec.\ref{Sec:DiscussConclusion} we draw some
concluding remarks about the modeling approach developed in the paper and about the design perspectives for the
Thin-TFET.

\section{Modeling of the tunneling transistor}\label{Sec:Model}

\subsection{Device concept and electrostatics}\label{Sec:DevElectro}

\begin{figure}[!ht]
\centering
\subfigure[]{
\includegraphics[width=0.7\textwidth]{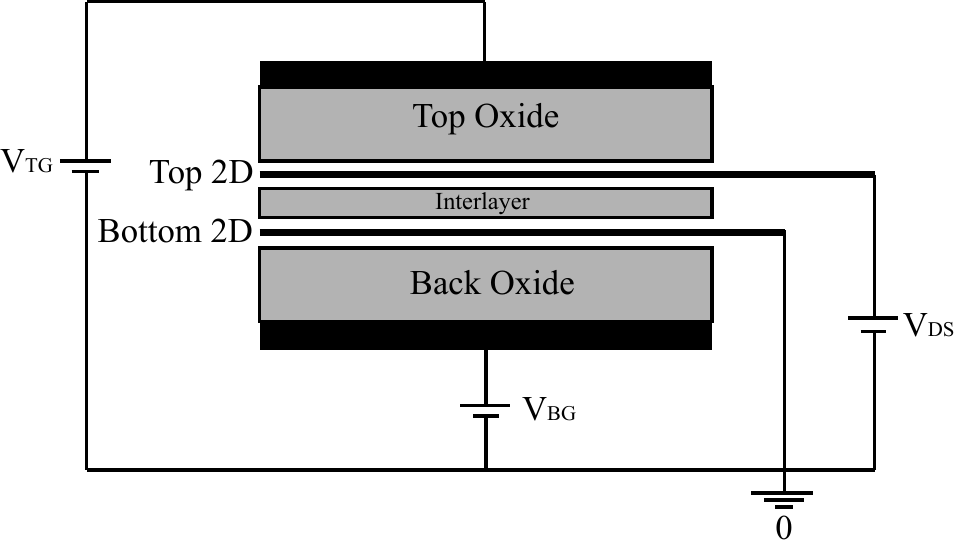}
}
\subfigure[]{
\includegraphics[width=0.7\textwidth]{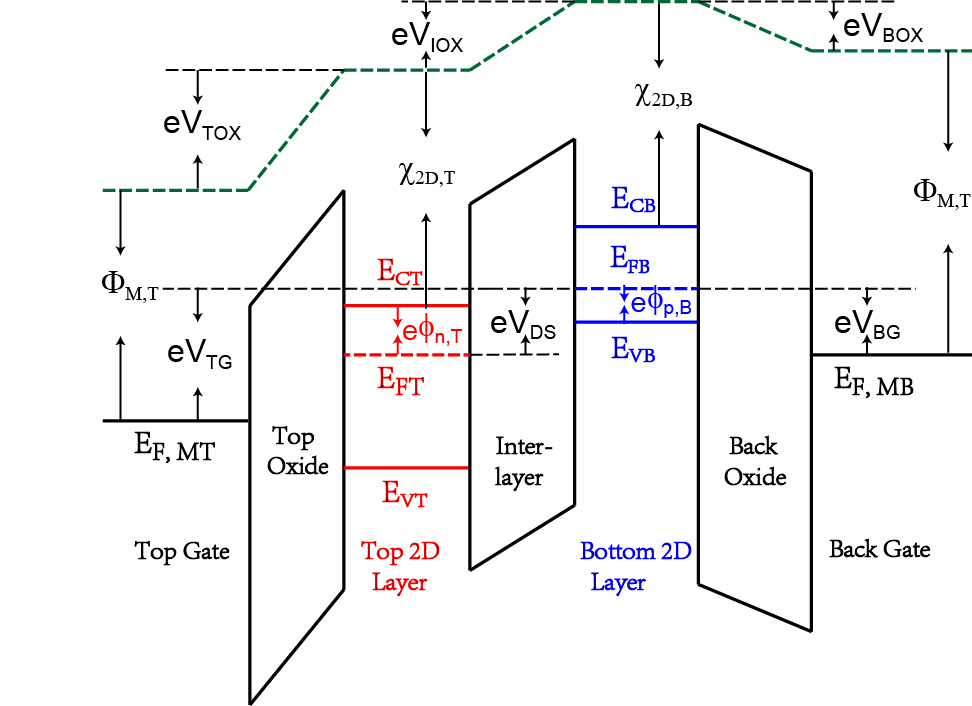}
} \caption[x]{ (a) Schematic device structure for the Thin-TFET, where V$_{TG}$, V$_{BG}$ and V$_{DS}$ are the top gate, bottom
gate and drain to source voltages; (b) sketch of the band diagram,
where $\Phi_{M,T}$, $\Phi_{M,B}$ are the work-functions and
E$_{F,MT}$, E$_{F,MB}$ the Fermi levels of the metal gates, while
$\chi_{2D,T}$, $\chi_{2D,B}$ are the electron affinities, E$_{FT}$,
E$_{FB}$ the Fermi levels, E$_{CT}$, E$_{CB}$ the conduction band
edges and E$_{VT}$, E$_{VB}$ the valence band edges respectively in the
top and bottom 2D layer. $V_{TOX}$, $V_{IOX}$ and $V_{BOX}$ are the
potential drops respectively across the top oxide, interlayer and bottom oxide.}\label{Fig:Device_BS}
\end{figure}

The device structure  and the corresponding band diagram are
sketched in Fig.\ref{Fig:Device_BS}, where the 2D materials are
assumed to be semiconductors with sizable energy bandgap, for example, transition-metal dichalcogenide (TMD) semiconductors without losing generality
\cite{mak2010atomically, wang2012electronics}.
Both the top 2D and the bottom 2D material is a monolayer and the
thickness of the 2D layers is neglected in the modeling of the electrostatics.

\begin{figure}[!ht]
\centering
\subfigure[]{
\includegraphics[width=0.43\textwidth]{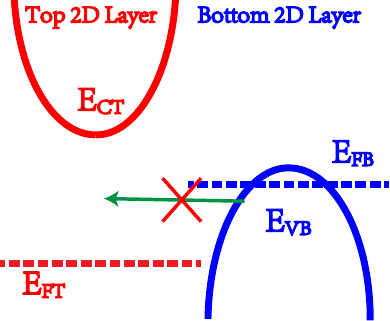}
}
\hspace{1em}
\subfigure[]{
\includegraphics[width=0.43\textwidth]{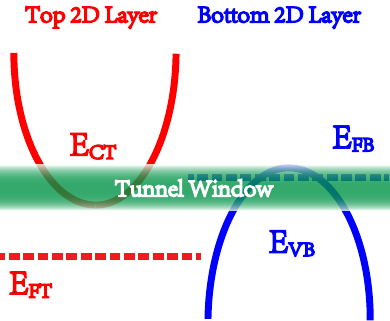}
} \caption[x]{ Sketch of the band alignments in a Thin-TFET between the top and
bottom 2D layer in: (a) OFF state and (b) ON state.
}\label{Fig:OFF/ON}
\end{figure}

The working principle of the tunneling transistor sketched in
Fig.\ref{Fig:Device_BS}(a) can be explained as follows. When the
conduction band edge E$_{CT}$ of the top 2D layer is higher than the
valence band edge E$_{VB}$ of the bottom 2D layer (see
Fig.\ref{Fig:OFF/ON}(a)), there are no states in the top layer to
which the electrons of the bottom layer can tunnel into. This
corresponds to the off state of the device. When E$_{CT}$
is pulled below E$_{VB}$ (see Fig.\ref{Fig:OFF/ON}(b)), a
tunneling window is formed and consequently an interlayer tunneling
can flow from the bottom to the top 2D material. The crossing and
uncrossing between the top layer conduction band and the bottom
layer valence band is governed by the gate voltages and it is
described by the electrostatics of the device.

To calculate the band alignment along the vertical direction of the intrinsic device in Fig.\ref{Fig:Device_BS} we write the Gauss law linking the sheet
charge in the 2D materials to the electric fields in the surrounding
insulating layers, which leads to
\begin{equation}
\begin{split}
\label{Eq:OneDP}
C_{TOX} V_{TOX}-C_{IOX} V_{IOX}=e(p_T-n_T+N_{D})\\
C_{BOX} V_{BOX}+C_{IOX} V_{IOX}=e(p_B-n_B+N_{A})
\end{split}
\end{equation}

\noindent where $C_{T(I, B)OX}$ is the capacitance per unit area of
top oxide (interlayer, bottom oxide) and $V_{T(I, B)OX}$ is the
potential drop across top oxide (interlayer, bottom oxide). The
potential drop across the oxides can be written in terms of the
external voltages $V_{TG}$, $V_{BG}$, $V_{DS}$ and of the energy
$e\phi_{n,T}=E_{CT}-E_{FT}$ and $e\phi_{p,T}=E_{FB}-E_{VB}$ defined in
Fig.\ref{Fig:Device_BS}(b) as
\begin{equation}
\begin{split}
\label{Eq:VoltageDrop}
e V_{TOX}=e V_{TG}  + e \phi_{n,T} - e V_{DS} + \chi_{2D, T} - \Phi_{M, T} \\
e V_{BOX}=e V_{BG} - e \phi_{p,B} + E_{GB} + \chi_{2D, B} + \Phi_{M, B} \\
e V_{IOX}=e V_{DS} - e \phi_{p,B} - e \phi_{n,T} + E_{GB} + \chi_{2D, B} - \chi_{2D, T}
\end{split}
\end{equation}

\noindent where $E_{FT}$, $E_{FB}$ are fermi levels of majority carriers in the top and bottom layer. $n_T$, $p_T$ are the electron and hole concentration
in the top layer, $n_B$, $p_B$ the concentrations in bottom layer,
$\chi_{2D, T}$, $\chi_{2D, B}$ are the electron affinities of the 2D
materials, $\Phi_T$, $\Phi_B$ the workfunctions of the top and back
gate and E$_{GB}$ is the energy gap in the bottom layer. Eq.
\ref{Eq:VoltageDrop} implicitly assumes that the majority carriers of the two 2D materials
are at thermodynamic equilibrium with their Fermi levels, with the
split of the Fermi levels set by the external voltages (i.e.
E$_{FB}$$-E_{FT}$$=$eV$_{DS}$), and the electrostatic potential
essentially constant in the 2D layers.

Since in our numerical calculations we shall employ a parabolic
effective mass approximation for the energy dispersion of the 2D
materials, as discussed more thoroughly in Sec.\ref{Sec:Results}, the carrier densities can be readily expressed as an analytic
function of $e\phi_{n,T}$ and $e\phi_{p,B}$ \cite{esseni2011nanoscale}
\begin{equation}
\label{Eq:Charge}
    n(p)=\frac{g_v m_c(m_v) k_B T}{\pi \hbar^2}\ln \left [  \exp \left ( -\frac{q \phi_{n,T}(\phi_{p,B})}{k_B T} \right )+1 \right ]
\end{equation}

\noindent where $g_v$ is the valley degeneracy.

When Eq.\ref{Eq:VoltageDrop} and Eq.\ref{Eq:Charge} are inserted in
Eq.\ref{Eq:OneDP}, we obtain two algebraic equations for $\phi_{n,T}$
and $\phi_{p,B}$ that can be solved numerically and describe the
electrostatics in a one dimensional section of the device.

\subsection{Transport model}\label{Sec:Transport}

In this section we develop a formalism to calculate the tunneling
current based on the transfer-Hamiltonian method
\cite{bardeen1961tunnelling, harrison1961tunneling, duke1969tunneling}, as also revisited recently for resonant
tunneling in graphene transistors
\cite{zhao2012symfet, feenstra2012single, britnell2013resonant}. We start
by writing the single particle elastic tunneling current as
\begin{equation}
\begin{split}
\label{Eq:TunnelCurrent1} I=g_v \frac{4 \pi e}{\hbar}
\displaystyle{\sum_{\kv_T, \kv_B}} |M(\kv_T, \kv_B)|^2
\delta (E_B(\kv_B)-E_T(\kv_T)) (f_B-f_T)\\
\end{split}
\end{equation}

\noindent where $e$ is the elementary charge, \kv$_B$, \kv$_T$ are
the \wavevs\ respectively in the bottom and top 2D material,
$E_B(\kv_B)$ $E_T(\kv_T)$ denote the corresponding energies, $f_B$
and $f_T$ are the Fermi occupation functions in the bottom and top
layer (depending respectively on E$_{FB}$ and E$_{FT}$, see
Fig.\ref{Fig:Device_BS}), and $g_v$ is the valley degeneracy.
The matrix element $M(\kv_T, \kv_B)$ expresses the transfer of
electrons between the two 2D layers is given by
\cite{britnell2013resonant}
\begin{equation}
\begin{split}
\label{Eq:MatrixElement1} M(\kv_T, \kv_B)=\displaystyle{\int_{A}}
d\rv \int dz \, \psi_{T,\kv_T}^{\dag}(\rv,z) \,
U_{sc}(\rv,z) \, \psi_{B,\kv_B} (\rv,z)\\
\end{split}
\end{equation}
\noindent where $\psi_{B,\kv_B}$ ($\psi_{T,\kv_T}$) is the electron
\wavef\ in the bottom (top) 2D layer and $U_{sc}(\rv,z)$ is the
perturbation potential in the interlayer region.

Eq.\ref{Eq:MatrixElement1} acknowledges the fact that in real
devices several physical mechanisms occurring in the interlayer
region can result in a relaxed conservation of the in plane \wavev\
\kv\ in the tunneling process.
We will return to the discussion of $U_{sc}(\rv,z)$ in this section.

To proceed in the calculation of $M(\kv_T, \kv_B)$ we write the
electron \wavef\ in the Bloch function form as
\begin{equation}
\label{Eq:Wavefunction}
\psi_{\kv}(\rv, z)=\frac{1}{\sqrt{N_C}} \,\me^ { i \kv \cdot \rv} \, u_{\kv}(\rv, z)\\
\end{equation}
\noindent where $u_{\kv}(\rv, z)$ is a periodic function of \rv\ and
$N_C$ is the number of unit cells in the overlapping area $A$ of the two 2D materials.
Eq.\ref{Eq:Wavefunction} assumes the following normalization condition:
\begin{equation}
\begin{split}
\label{Eq:Normalization} \int_{\Omega_C} d\rhov \int_z dz
|u_{\kv}(\rhov, z)|^2=1
\end{split}
\end{equation}
\noindent where \rhov\ is the in-plane abscissa in the unit cell
area $\Omega_C$ and $A$=$N_C$$\Omega_C$.

The \wavef\ $\psi_{\kv}(\rv, z)$ is assumed to decay exponentially
in the interlayer region with a decay constant $\kappa$
\cite{feenstra2012single,britnell2013resonant}; such a $z$ dependence is absorbed in $u_{\kv}(\rv, z)$ and
we do not need to make it explicit in our derivations. It should be
noticed that absorbing the exponential decay in $u_{\kv}(\rv, z)$
recognizes the fact that in the interlayer region the \rv\ dependence of
the \wavef\ changes with $z$. In fact, as already discussed
\cite{feenstra2012single}, while the $u_{\kv}(\rv, z)$ are localized
around the basis atoms in the two 2D layers, these functions are
expected to spread out while they decay in the interlayer region, so
that the \rv\ dependence becomes weaker when moving farther from the
2D layers.

To continue in the calculation of $M(\kv_T, \kv_B)$ we let the
scattering potential in the interlayer region be separable in the
form \cite{britnell2013resonant}
\begin{equation}
U_{sc}(\rv, z)= V_B(z) \, F_{L}(\rv)
\label{Eq:UscSeparation}
\end{equation}
where $F_{L}(\rv)$ is the in-plane fluctuation of the scattering
potential, which is essentially responsible for the relaxation of
momentum conservation in the tunneling process.

By substituting Eqs.\ref{Eq:Wavefunction} and \ref{Eq:UscSeparation}
in Eq.\ref{Eq:MatrixElement1} and writing $\rv$$=$$\rv_j$$+$$\rhov$,
where $\rv_j$ is a direct lattice vector and $\rhov$ is the in-plane
position inside each unit cell, we obtain
\begin{eqnarray}
\label{Eq:MatrixElement2} M(\kv_T, \kv_B) & = &\frac{1}{N_C}
\sum_{j=1}^{N_C}  \me ^{i (\kv_B-\kv_T) \cdot \rv_j }
\int_{\Omega_C} d\rhov \int dz \, \me ^{i (\kv_B-\kv_T) \cdot \rhov}
\times \, \nonumber \\ & \times & u_{T,\kv_T}^{\dag}(\rv_j+\rhov, z)
\, F_{L}(\rv_j+\rhov) V_B(z) \, u_{B,\kv_B}(\rv_j+\rhov, z)
\end{eqnarray}

\noindent We now assume that F$_L$(\rv) corresponds to relatively
long range fluctuations so that it can be taken as approximately
constant inside a unit cell, and that, furthermore, the top and
bottom 2D layer have the same lattice constant, hence the Bloch
functions $u_{T, \kv_T}$ and $u_{B, \kv_B}$ have the same periodicity
in the \rv\ plane.
Moreover, for the time being we consider that the conduction band
minimum in the top layer and the valence band maximum in the bottom
layer are at the same point of the 2D Brillouin zone, so that
$\qv$$=$$\kv_B$$-$$\kv_T$ is small compared to the size of the
Brillouin zone and $\me^{i \qv \cdot \rhov }$ is approximately 1.0
inside a unit cell. These considerations and approximations allow us
to rewrite Eq.\ref{Eq:MatrixElement2} as
\begin{equation}
\begin{split}
\label{Eq:MatrixElement3} M(\kv_T, \kv_B) \simeq \frac{1}{N_C}
\sum_{j=1}^{N_C} \me ^{i \qv \cdot \rv_j } F_{L}(\rv_j)
\displaystyle{\int_{\Omega_C}}d\rhov \int dz \,
u_{T,\kv_T}^{\dag}(\rhov,z) \, V_B(z) \, u_{B,\kv_B}(\rhov, z)
\end{split}
\end{equation}
where the integral in the unit cell has been written for
\rv$_j$$=$\zerov\ because it is independent of the unit cell.

Consistently with the assumption that \kv$_B$ and \kv$_T$ are small
compared to the size of the Brillouin zone, in
Eq.\ref{Eq:MatrixElement3} we neglect the \kv$_B$ (\kv$_T$)
dependence of u$_{B,\kv_B}$ (u$_{T,\kv_T}$) and simply set
u$_{T,\kv_T}(\rhov, z)$$\approx$u$_{0T}(\rhov, z)$,
u$_{B,\kv_B}(\rhov, z)$$\approx$u$_{0B}(\rhov, z)$, where
u$_{0T}(\rhov, z)$ and u$_{0B}(\rhov, z)$ are the periodic parts of
the Bloch function at the band edges, which is the simplification
typically employed in the effective mass approximation approach
\cite{esseni2011nanoscale}. By recalling that the u$_{0B}$ and
u$_{0T}$ retain the exponential decay of the \wavefs\ in the
interlayer region with a decay constant $\kappa$, we now write
\begin{equation}
\label{Eq:MB0_def}
\displaystyle{\int_{\Omega_C}}d\rhov \int dz \,
u_{0T}^{\dag}(\rhov,z) \, V_B(z) \, u_{0B}(\rhov, z) \simeq M_{B0}
\, \me ^{- \kappa T_{IL}}
\end{equation}
\noindent where T$_{IL}$ is the interlayer thickness and $M_{B0}$ is
a \kv\ independent matrix element that will remain a prefactor in
the final expression for the tunneling current.
Since F$_L(\rv)$ has been assumed a slowly varying function over a
unit cell, then the sum over the unit cells in
Eq.\ref{Eq:MatrixElement3} can be rewritten as a normalized integral
over the tunneling area
\begin{equation}
\begin{split}
\label{Eq:FL_Fourier} \frac{1}{\Omega_c \, N_C} \sum_{j=1}^{N_C}
\Omega_c \, \me ^{i \qv \cdot \rv_j } F_{L}(\rv_j) \simeq
\frac{1}{A} \displaystyle{\int_{A}} \, \me ^{i \qv \cdot \rv }
F_{L}(\rv) d\rv
\end{split}
\end{equation}

By introducing Eq.\ref{Eq:MB0_def} and \ref{Eq:FL_Fourier} in
Eq.\ref{Eq:MatrixElement3} we can finally express the squared matrix
element as
\begin{equation}
\begin{split}
\label{Eq:MatrixElement4} |M(\kv_T, \kv_B)|^2 \simeq
\frac{|M_{B0}|^2 \, S_{F} (\qv)}{A} \, \me ^{- 2\kappa T_{IL}}
\end{split}
\end{equation}
\noindent where $\qv$$=$$\kv_B$$-$$\kv_T$ and S$_{F}$(\qv) is the
power spectrum of the random fluctuation described by $F_{L}(\rv)$,
which is defined as
\cite{esseni2011nanoscale}
\begin{equation}
\begin{split}
\label{Eq:Spectrum_DEF}
S_{F}(\qv) = \frac{1}{A} \left |
\displaystyle{\int_{A}} \, \me ^{i \qv \cdot \rv } F_{L}(\rv) d\rv \right
|^2
\end{split}
\end{equation}

\noindent By substituting Eq.\ref{Eq:MatrixElement4} in
Eq.\ref{Eq:TunnelCurrent1} and then converting the sums over \kv$_B$
and \kv$_T$ to integrals we obtain
\begin{equation}
\begin{split}
\label{Eq:TunnelCurrent2} I=\frac{g_v e \, |M_{B0}|^2 \, A}{4 \pi^3
\hbar} \me ^{- 2 \kappa T_{IL}} \displaystyle{\int_{\kv_T}}
\displaystyle{\int_{\kv_B}} d\kv_T \, d\kv_B \, S_{F}(\qv) \, \delta
(E_B(\kv_B)-E_T(\kv_T)) \, (f_B-f_T)
\end{split}
\end{equation}

\noindent Before we proceed with some important integrations of the
basic model that will be discussed in Secs.\ref{Sec:broadening} and
\ref{Sec:misalignment}, a few comments about the results obtained so
far are in order below.

According to Eq.\ref{Eq:TunnelCurrent2} the current is proportional
to the squared matrix element $|M_{B0}|^2$ defined in
Eq.\ref{Eq:MB0_def} and decreases exponentially with the thickness interlayer
$T_{IL}$ according to the decay constant $\kappa$
of the \wavefs. Attempting to derive a quantitative expression for
$M_{B0}$ is admittedly very difficult, in fact it
is difficult to determine how the periodic functions u$_{0T}(\rhov,
z)$ and u$_{0B}(\rhov, z)$ spread out when they decay in the barrier
region and, furthermore, it is not even perfectly clear what
potential energy or Hamiltonian should be used to describe the
barrier region itself, which is an issue already recognized and
thoroughly discussed in the literature since a long time
\cite{duke1969tunneling}. Our model essentially circumvents these
difficulties by resorting to the semi-empirical formulation of the
matrix element given by Eq.\ref{Eq:MB0_def}, where $M_{B0}$ is left
as a parameter to be determined and discussed by comparing to
experiments.

It is also worth noting that in our calculations we have not
explicitly discussed the effect of spin-orbit interaction in the
bandstructure of 2D materials, even if giant spin-orbit couplings
have been reported in 2D transition-metal dichalcogenides
\cite{zhu2011giant}. If the energy separations between the spin-up
and spin-down bands are large, then the spin degeneracy in current
calculations should be one instead of two, which would affect the
current magnitude but not its dependence on the gate bias.
Our calculations neglected also the possible modifications of band
structure in the TMD materials produced by the vertical electrical
field, in fact we believe that in our device the electrical field in the 2D layers
is not strong enough to make such effects significant
\cite{ramasubramaniam2011tunable}.

The decay constant $\kappa$ in the interlayer region may be
estimated from the electron affinity difference between the 2D
layers and the interlayer material \cite{feenstra2012single}.
Moreover, according to Eq.\ref{Eq:TunnelCurrent2} the constant
$\kappa$ determines the dependence of the current on $T_{IL}$, so
that $\kappa$ may be extracted by comparing to experiments
discussing such a dependence, which, for example, have been recently
reported for the interlayer tunneling current in a graphene-$h$BN system
\cite{britnell2012atomically}.

As for the spectrum S$_{F}$(\qv) of the scattering potential, in our
calculations we utilize
\begin{equation}
\label{Eq:Scr} S_R(q)=
\frac{\pi L_C^2}{(1+q^2 L_C^2 / 2)^{3/2}}
\end{equation}
where $L_C$ is the correlation length, which in our derivations has
been assumed large compared to the size of a unit cell.
Eq.\ref{Eq:Scr} is consistent with an exponential form for the
autocorrelation function of F$_L$(\rv) \cite{esseni2011nanoscale},
and a similar $\qv$ dependence has been recently employed to
reproduce the experimentally observed line-width of the resonance
region in graphene interlayer tunneling transistors
\cite{britnell2013resonant}. Such a functional form can be
representative of phonon assisted tunneling, short-range disorder\cite{li2011theory}, charged
impurities\cite{yan2011correlated}
or Moir\'e patterns that have been observed, for instance, at the graphene-$h$BN interface\cite{yankowitz2012emergence,xue2011scanning,decker2011local}.
We will see in Sec.\ref{Sec:AnalyticResults} that the $L_C$ has an
influence on the gate voltage dependent current, which has a
neat physical interpretation,
hence a comparison to experimental data will be very informative for an estimate of $L_C$.

\subsection{Effects of energy broadening}\label{Sec:broadening}

According to Eq.\ref{Eq:TunnelCurrent1} and Eq.\ref{Eq:TunnelCurrent2} the tunneling current is
simply zero when there is no energy overlap between the conduction
band in the top layer and the valence band in the bottom layer, that
is for $E_{CT}$$>$$E_{VB}$. In a real device, however, the 2D
materials will inevitably have phonons, disorder, host impurities in the 2D
layer and be affected by the background impurities in the
surrounding materials, so that a finite broadening of the energy
levels is expected to occur because of the statistical potential
fluctuations superimposed to the ideal crystal structure
\cite{van1991generalized}.
The energy broadening in 3D semiconductors is known to lead to a
tail of the density of states (DoS) in the gap region, that has been
also observed in optical absorption measurements and denoted Urbach
tail \cite{urbach1953long,cody1992urbach}. It is thus
expected that the finite energy broadening will be a fundamental
limit to the abruptness of the turn on characteristic attainable
with the devices of this work, hence it is important to include this
effect in our model.

Energy broadening in the 2D systems can stem from the interaction
with randomly distributed impurities and disorder in the 2D layer or
in the surrounding materials
\cite{kane1963thomas,van1991generalized,sarma1982thomas},
by scattering events induced by the interfaces
\cite{knabchen1995self}, as well as by other scattering sources. We
recognize the fact that a detailed description of the energy
broadening is exceedingly complicated due to the many-body and
statistical fluctuation effects \cite{ghazali1985disorder}, and thus
resort to a relatively simple semi-classical treatment\cite{kane1963thomas}$^,$
\cite{van1991generalized}.
We start by recalling that the density of states $\rho_0(E)$ for a
2D layer with no energy broadening is
\begin{equation}
\begin{split}
\label{Eq:DOSdef} \rho_0(E)=\frac{g_s g_v}{4\pi^2}
\displaystyle{\int_\kv} d\kv \, \delta \left[E-E(\kv)\right]
\end{split}
\end{equation}
where E(\kv) denotes the energy relation with no broadening and $g_s$, $g_v$ are spin and valley degeneracy. In the
presence of a randomly fluctuating potential V(\rv), instead, the
DoS can be written as \cite{kane1963thomas, van1991generalized}
\begin{equation}
\begin{split}
\label{Eq:DOSBroadening2}
\rho(E)&=\int_{0}^\infty dv \, \rho_0(v) P_v(E-v)\\
&=\frac{g_s g_v}{4\pi^2}  \int_\kv d\kv \left [ \int_{0}^\infty dv \,
\delta\left [v - E(\kv) \right ] P_v (E-v) \right] \\
&=\frac{g_s g_v}{4\pi^2} \int_\kv d\kv  \, P_v\left[E- E(\kv)\right]
\end{split}
\end{equation}
where P$_v$(v) is the distribution function for V(\rv) (to be
further discussed below), and we have used the $\rho_0(E)$
definition in Eq.\ref{Eq:DOSdef} to go from the first to the second
equality.

Comparing Eq.\ref{Eq:DOSBroadening2} to Eq.\ref{Eq:DOSdef}, we see
that the $\rho$(E) of the system in the presence of broadening can
be calculated by substituting the Dirac function in
Eq.\ref{Eq:DOSdef} with a finite width function P$_v$(v), which is
the distribution function of V(\rv) and it is thus normalized to
one.

In order to include the energy broadening in our current
calculations, we rewrite the tunneling rate in
Eq.\ref{Eq:TunnelCurrent1} as
\begin{equation}
\begin{split}
\label{Eq:TunnelProb}
\frac{1}{\tau_{\kv_T,\kv_B}}&=\frac{2\pi}{\hbar}|M(\kv_T, \kv_B)|^2 \delta \left [E_T(\kv_T)-E_B(\kv_B)\right ]\\
&=\frac{2\pi}{\hbar}|M(\kv_T, \kv_B)|^2 \int_{-\infty}^{\infty} dE  \delta \left [E- E_T(\kv_T) \right ] \delta \left [ E- E_B(\kv_B)\right ] \\
\end{split}
\end{equation}
and note that, consistently with Eq.\ref{Eq:DOSBroadening2}, the
energy broadening can be included in the current calculation by
substituting $\delta [ E- E(\kv)]$ with $P_v[ E- E(\kv)]$. By
doing so the tunneling rate becomes
\begin{equation}
\begin{split}
\label{Eq:TunnelProb2}
\frac{1}{\tau_{\kv_T,\kv_B}}& \simeq
\frac{2\pi}{\hbar}|M(\kv_T, \kv_B)|^2  S_E(E_T(\kv_T)-E_B(\kv_B))
\end{split}
\end{equation}
where we have introduced an energy broadening spectrum $S_E$ that is
defined as
\begin{equation}
\begin{split}
\label{Eq:TunnelProb3} S_E(E_T(\kv_T)-E_B(\kv_B)) =
\int_{-\infty}^{\infty} dE P_{vT} \left [ E- E_T(\kv_T) \right ]
P_{vB} \left [E-E_B(\kv_B) \right ]
\end{split}
\end{equation}
where $P_{vT}$ and $P_{vB}$ is the potential distribution function due to the presence of randomly fluctuating potential $V(\rv)$ in the top and the bottom layer, respectively.

On the basis of Eq.\ref{Eq:TunnelProb2}, in our model for the
tunneling current we accounted for the energy broadening by using in
all numerical calculations the broadening spectrum
$S_E(E_T(\kv_T)$$-$$E_B(\kv_B))$ defined in Eq.\ref{Eq:TunnelProb3}
in place of $\delta [E_T(\kv_T)$$-$$E_B(\kv_B)]$.
More precisely  we used a Gaussian potential distribution for both
the top and the bottom layer
\begin{equation}
\begin{split}
\label{Eq:Distri.}
P_v(E-E_{\kv0})=\frac{1}{\sqrt{\pi}\sigma} \me ^{-(E-E_{\kv0})^2/\sigma^2}\\
\end{split}
\end{equation}
which has been derived by Evan O.Kane for a broadening induced by
randomly distributed impurities \cite{kane1963thomas}, in which case
$\sigma$ can be expressed in terms of the average impurity
concentration.

Quite interestingly, for the Gaussian spectrum in
Eq.\ref{Eq:Distri.} the overall broadening spectrum $S_E$ defined in
Eq.\ref{Eq:TunnelProb3} can be calculated analytically and reads
\begin{equation}
\begin{split}
\label{Eq:SE}
S_E(E_T(\kv_T)-E_B(\kv_B)) = \frac{1}{\sqrt\pi(\sigma_T^2+\sigma_B^2)} \me^{-(E_T(\kv_T)-E_B(\kv_B))^2/\sigma^2}\\
\hspace{5mm} .
\end{split}
\end{equation}

\noindent Hence also S$_E$ has a Gaussian spectrum,
where $\sigma_T$ and $\sigma_B$ are the broadening energies for the
top and bottom 2D layer, respectively.

\subsection{Rotational misalignment and tunneling between inequivalent extrema}
\label{Sec:misalignment}

\begin{figure}[!ht]
\centering
\includegraphics[width=0.5\textwidth]{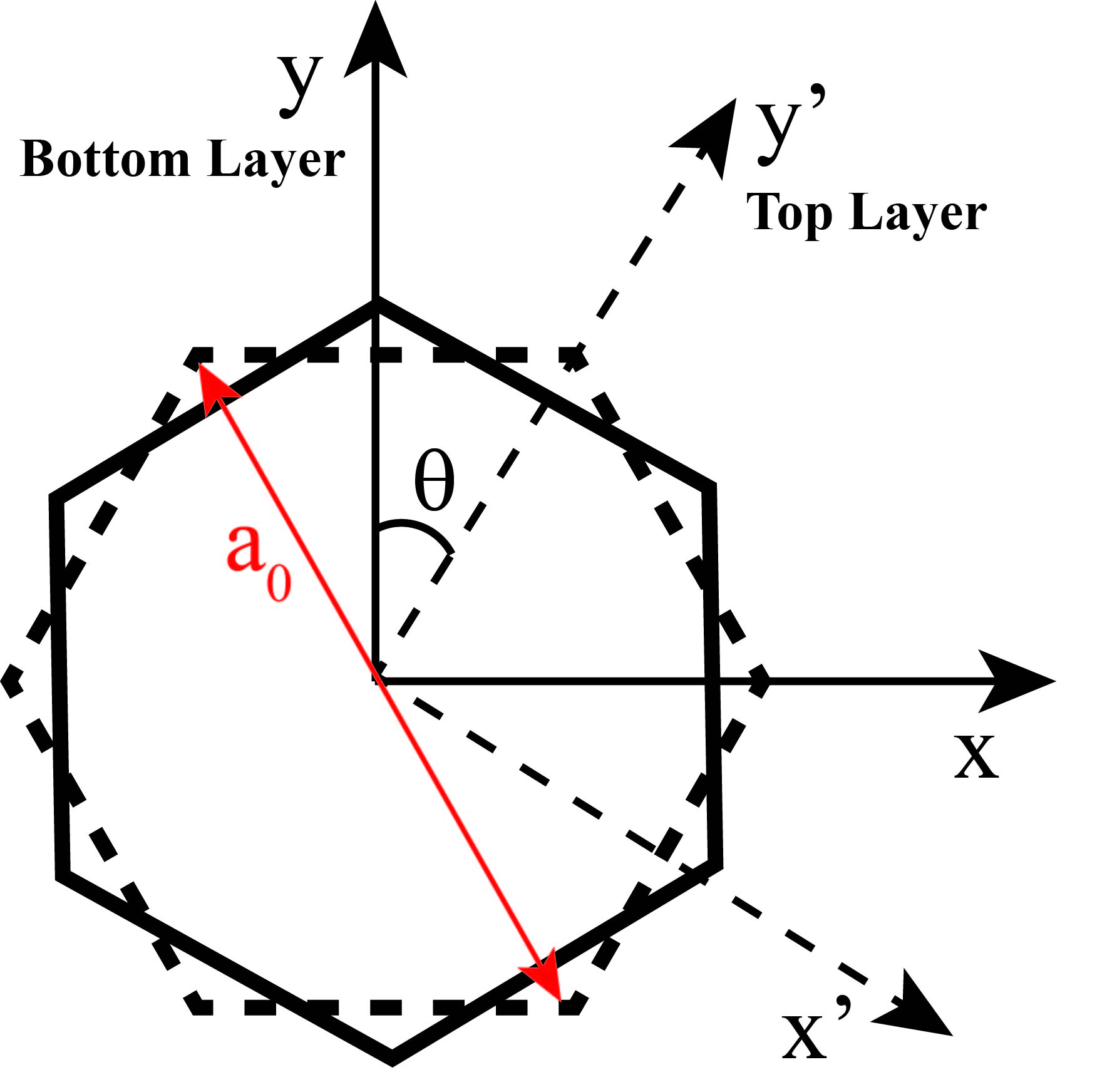}
\caption[x]{ Sketch of a possible rotational misalignment between the top and bottom
2D layer, x-y is the reference coordinate for the bottom 2D layer and
x'-y' is the reference coordinate for the top 2D layer. $\theta$ is the
rotational misalignment angle. We assume the top layer and the bottom layer have the same lattice
constant $a_0$. }\label{Fig:Misalignment}
\end{figure}

The derivations in Sec.\ref{Sec:Transport} assumed that there is a
perfect rotational alignment between the top and the bottom layer
and that the tunneling occurs between equivalent extrema in the
Brillouin zone, that is tunneling from a $K$ to a $K$ extremum (or
from $K'$ to $K'$ extremum). We now denote by $\theta$ the angle
expressing a possible rotational misalignment between the two 2D
layers (see Fig.\ref{Fig:Misalignment}), and still assume that the
top 2D crystal has the same lattice constant $a_0$ as the bottom 2D
crystal. The principal coordinate system is taken as the crystal
coordinate system in the bottom layer, and we denote with \rvp,
\kvp\ the position and wave vectors in the crystal coordinate system
of the top layer (with \rv, \kv\ being the vectors in the principal
coordinate system). The \wavef\ in the top layer has the form given
in Eq.\ref{Eq:Wavefunction} in terms of \rvp, \kvp, hence in order
to calculate the matrix element in the principal coordinate system
we start by writing \rvp$=$$\hat{R}_{B \rightarrow T}$\rv,
\kvp$=$$\hat{R}_{B \rightarrow T}$\kv, where $\hat{R}_{B \rightarrow
T}$ is the rotation matrix from the bottom to the top coordinate
system, with $\hat{R}_{T \rightarrow B}$$=$$[\hat{R}_{B \rightarrow
T}]^T$ being the matrix going from the top to the bottom coordinate
system and $M^T$ denoting the transpose of the matrix $M$. The
rotation matrix can be written as
\begin{equation}
\label{Eq:RotationMatrix} \hat{R}_{T\rightarrow B} = \left(
\begin{array}{ccc}
cos\theta & -sin\theta \\
sin\theta & cos\theta \end{array} \right)
\end{equation}
in terms of the rotational misalignment angle $\theta$.

Consistently with Sec.\ref{Sec:Transport} we set u$_{T,\kv_T}(\rvp,
z)$$\approx$u$_{0T}(\rvp,z)$, u$_{B,\kv_B}(\rv,
z)$$\approx$u$_{0B}(\rv,z)$, where u$_{0T}(\rvp,z)$,
u$_{0B}(\rv,z)$ are the periodic part of the Bloch function
respectively at the band edge in the top and bottom layer. We then
denote with $\Kv_{0T}$ the \wavev\ at the conduction band edge in
the top layer (expressed in the top layer coordinate system), and
with $\Kv_{0B}$ the \wavev\ at the valence band edge in the bottom
layer (expressed in the principal coordinate system); the
derivations in this section account for the fact that $\Kv_{0T}$ and
$\Kv_{0B}$ may be inequivalent extrema (i.e.
$\Kv_{0T}$$\neq$$\Kv_{0B}$).

By expressing \rvp\ and \kvp\ in the principal coordinate system we
can essentially follow the derivations in Sec.\ref{Sec:Transport}
and write the matrix element as
\begin{eqnarray}
\label{Eq:MatrixElementMisalign} M(\kv_T, \kv_B) & \simeq &
\frac{1}{N_C} \sum_{j=1}^{N_C} \me ^{i (\qv+\Qv_D) \cdot \rv_j }
F_{L}(\rv_j) \times \nonumber \\
& \times & \int_{\Omega_{C}} d\rv \int dz \,
u_{0T}^{\dag}(\hat{R}_{B\rightarrow T}(\rv_j+\rhov), z) \, V_{B}(z)
\, u_{0B}(\rv_j+\rhov, z)
\end{eqnarray}
where \qv$=$$($\kv$_B$$-$\kv$_T$$)$ and we have introduced the
vector
\begin{equation}
\begin{split}
\label{Eq:Misalign} \Qv_D=\Kv_{0B}-\hat{R}_{T\rightarrow B}\Kv_{0T}
\end{split}
\end{equation}

\noindent Eq.\ref{Eq:MatrixElementMisalign} is an extension of
Eq.\ref{Eq:MatrixElement3} that accounts for a possible rotational
misalignment between the 2D layers and describes also the tunneling
between inequivalent extrema. The vector $\Qv_D$ is zero only for
tunneling between equivalent extrema (i.e. $\Kv_{0B}$$=$$\Kv_{0T}$)
and for a perfect rotational alignment (i.e. $\theta$$=$0). Considering a case where all extrema are at the $\Kv$ point, 
we have $|$$\Kv_{0B}$$|$$=$$|$$\Kv_{0T}$$|$$=$4$\pi$$/$3$a_0$, 
then for $\Kv_{0B}$$=$$\Kv_{0T}$ the magnitude of $\Qv_D$ is simply given by
$Q_D$$=$$(8\pi/3a_0)\sin(\theta/2)$ \cite{feenstra2012single}.

One significant difference in Eq.\ref{Eq:MatrixElementMisalign}
compared to Eq.\ref{Eq:MatrixElement3} is that, in the presence of
rotational misalignment, the top layer Bloch function
$u_{0T}(\hat{R}_{B\rightarrow T}\rv, z)$ has a different periodicity
in the principal coordinate system from the bottom layer
$u_{0B}(\rv, z)$. Consequently the integral over the unit cells of
the bottom 2D layer is not the same in all unit cells, so that the
derivations going from Eq.\ref{Eq:MatrixElement3} to
Eq.\ref{Eq:TunnelCurrent2} should be rewritten accounting for a
matrix element $M_{B0,j}$ depending on the unit cell $j$. Such an
$M_{B0,j}$ could be formally included in the calculations by
defining a new scattering spectrum that includes not only the
inherently random fluctuations of the potential $F_L(\rv)$, but also
the cell to cell variations of the matrix element $M_{B0,j}$.
A second important difference of Eq.\ref{Eq:MatrixElementMisalign}
compared to Eq.\ref{Eq:MatrixElement3} lies in the presence of
\Qv$_D$ in the exponential term multiplying F$_L(\rv_j)$.

For the case of tunneling between inequivalent extrema and with a
negligible rotational misalignment (i.e. $\theta$$\simeq$0),
Eq.\ref{Eq:Misalign} gives \Qv$_D$$=$\Kv$_{0B}$$-$\Kv$_{0T}$ and the current can be expressed as in
Eq.\ref{Eq:TunnelCurrent2} but with the scattering spectrum
evaluated at $|$\qv$+$\Qv$_D$$|$. Since in this case the magnitude
of \Qv$_D$ is comparable to the size of the Brillouin zone, the
tunneling between inequivalent extrema is expected to be
substantially suppressed if the correlation length $L_c$ of the
scattering spectrum S$_R$(q) is much larger than the lattice
constant, as it has been assumed in all the derivations.

Quite interestingly, the derivations in this section suggest that a
possible rotational misalignment is expected to affect the absolute
value of the tunneling current but not to change significantly its
dependence on the terminal voltages.

From a technological viewpoint, if the stack of the 2D materials is
obtained using a dry transfer method the rotational misalignment
appears inevitable \cite{britnell2013resonant, britnell2012field}. Experimental results have shown that,
when the stack of 2D materials is obtained by growing the one
material on top of the other,
the top 2D and bottom 2D layer can have a fairly good
angular alignment \cite{tiefenbacher2000moire, koma1999van}.

\subsection{An analytical approximation for the tunneling current}
\label{Sec:AnalyticResults}

The numerical calculations for the tunneling current obtained with
the model derived in Secs.\ref{Sec:Transport} and
\ref{Sec:broadening} will be presented in Sec.\ref{Sec:Results},
while in this section we discuss an analytical, approximated
expression for the tunneling current which is mainly useful to gain
an insight about the main physical and material parameters affecting
the current versus voltage characteristic of the Thin-TFET. In order to derive an
analytical current expression we start by assuming a parabolic
energy relation and write
\begin{equation}
\label{Eq:ParabBands} E_{VB}(\kv_B) = E_{VB} - \frac{\hbar^2
k_B^2}{2m_v} \hspace{10mm}  E_{CT}(\kv_T) = E_{CT} + \frac{\hbar^2
k_T^2}{2m_c}
\end{equation}
\noindent where $E_{VB}(\kv_B)$, $E_{CT}(\kv_T)$ are the energy relation
respectively in the bottom layer valence band and top layer
conduction band and $m_v$, $m_c$ the corresponding effective masses.

In the analytical derivations we neglect the energy broadening and
start from Eq.\ref{Eq:TunnelCurrent2}, so that the model is
essentially valid only in the on-state of the device, that is for
$E_{CT}$$<$$E_{VB}$.

We now focus on the integral over \kv$_B$ and \kv$_T$ in
Eq.\ref{Eq:TunnelCurrent2} and first introduce the polar coordinates
\kv$_B$$=$$($k$_B$,$\theta_B$$)$, \kv$_T$$=$$($ k$_T$,$\theta_T$$)$,
and then use Eq.\ref{Eq:ParabBands} to convert the integrals over
k$_B$, k$_T$ to integrals over respectively E$_B$, E$_T$, which
leads to
\begin{eqnarray}
\label{Eq:Analytic1} & & I \propto \displaystyle{\int_{\kv_T}}
\displaystyle{\int_{\kv_B}} d\kv_T \, d\kv_B \, S_{F}(q) \, \delta
(E_B(\kv_B)-E_T(\kv_T)) \, (f_B-f_T) \\ \nonumber
& =& \frac{m_c m_v}{\hbar^4} \displaystyle{\int_{0}^{2\pi}}
\hspace{-2mm} d\theta_B \displaystyle{\int_{0}^{2\pi}} \hspace{-2mm}
d\theta_T \displaystyle{\int_{E_{CT}}^{\infty}} \hspace{-2mm} dE_T
\displaystyle{\int_{-\infty}^{E_{VB}}} \hspace{-2mm} dE_B \,
S_{F}(q) \, \delta (E_B-E_T) \, (f_B-f_T)
\end{eqnarray}
where the spectrum $S_{F}$(q) is given by Eq.\ref{Eq:Scr} and thus
depends only on the magnitude $q$ of $\qv$$=$$\kv_B$$-$$\kv_T$.
\noindent Assuming $E_{CT}$$<$$E_{VB}$, the Dirac function reduces
one of the integrals over the energies and sets $E$$=$$E_B$$=$$E_T$,
furthermore the magnitude of $\qv$$=$$\kv_B$$-$$\kv_T$ depends only
on the angle $\theta$$=$$\theta_B$$-$$\theta_T$, so that
Eq.\ref{Eq:Analytic1} simplifies to
\begin{eqnarray}
I \propto \frac{m_c m_v (2\pi)}{\hbar^4} \displaystyle{\int_{0}^{2\pi}}
\hspace{-2mm} d\theta \displaystyle{\int_{E_{CT}}^{E_{VB}}}
\hspace{-2mm} dE \, S_{F}(q) \, (f_B-f_T)
\label{Eq:Analytic2}
\end{eqnarray}

\noindent In the on-state condition (i.e. for $E_{CT}$$<$$E_{VB}$),
the zero Kelvin approximation for the Fermi-Dirac occupation
functions f$_B$, f$_T$ can be introduced to further simplify
Eq.\ref{Eq:Analytic2} to
\begin{eqnarray}
I \propto \frac{m_c m_v (2\pi)}{\hbar^4} \displaystyle{\int_{0}^{2\pi}}
\hspace{-2mm} d\theta \displaystyle{\int_{E_{min}}^{E_{max}}}
\hspace{-2mm} dE \, S_{F}(q)
\label{Eq:Analytic3}
\end{eqnarray}
where $E_{min}$$=$max\{$E_{CT},E_{FT}$\},
$E_{max}$$=$min\{$E_{VB},E_{FB}$\} define the tunneling window
$[E_{max}-E_{min}]$.

The evaluation of Eq.\ref{Eq:Analytic3} requires to express $q$ as a
function of the energy $E$ inside the tunneling window and of the
angle $\theta$ between \kv$_B$ and \kv$_T$. By recalling
$q^2$$=$$k_B^2$$+$$k_T^2$$-$$2k_Bk_T \cos(\theta)$, we can
use Eq.\ref{Eq:ParabBands} to write
\begin{equation}
q^2 = \frac{2m_v}{\hbar^2}(E_{VB}-E) +
\frac{2m_c}{\hbar^2}(E-E_{CT}) - \frac{4 \sqrt{m_c
m_v}}{\hbar^2}\sqrt{ (E_{VB}-E)(E-E_{CT})} \,
\cos(\theta)\label{Eq:q_E_Theta}
\end{equation}
with $E$$=$$E_B$$=$$E_T$.
When Eq.\ref{Eq:q_E_Theta} is substituted in the spectrum S$_{F}$(q)
the resulting integrals over $E$ and $\theta$ in
Eq.\ref{Eq:Analytic3} cannot be evaluated analytically. Therefore to
proceed further we now examine the maximum value taken by $q^2$. The
$\theta$ value leading to the largest $q^2$ is $\theta$$=$$\pi$, and
the resulting $q^2$ expression can be further maximized with respect
to the energy $E$ varying in the tunneling window. The energy
leading to maximum $q^2$ is
\begin{equation}
E_{M} = \frac{E_{CT} + (m_c/m_v)E_{VB}}{1+(m_c/m_v)}
\label{Eq:EmaxQ}
\end{equation}
and the corresponding $q^2_M$ is
\begin{equation}
q^2_{M} = \frac{2(m_c+m_v)(E_{VB}-E_{CT})}{\hbar^2} \label{Eq:Qmax}
\end{equation}

When neither the top nor the bottom layer are degenerately doped the
tunneling window is given by E$_{min}$$=$E$_{CT}$ and
E$_{max}$$=$E$_{VB}$, in which case the E$_{M}$ defined in
Eq.\ref{Eq:EmaxQ} belongs to the tunneling window and the maximum
value of q$^2$ is given by Eq.\ref{Eq:Qmax}. If either the top or
the bottom layer is degenerately doped the Fermi levels become the edges of
the tunneling window and the maximum value of q$^2$ may be smaller
than in Eq.\ref{Eq:Qmax}.

A drastic simplification in the evaluation of Eq.\ref{Eq:Analytic3}
is obtained for $q^2_{M}$$\ll$1$/$$L_c^2$, in which case
Eq.\ref{Eq:Scr} returns to S$_{F}$(q)$\approx$$\pi L_c^2$, so that by
substituting S$_{F}$(q) in Eq.\ref{Eq:Analytic2} and then in
Eq.\ref{Eq:TunnelCurrent2} the expression for the current simplifies
to
\begin{equation}
I \simeq \frac{e g_v A (m_c m_v)}{\hbar^5} |M_{B0}|^2 \, \me ^{- 2
\kappa T_{IL}} \,  L_c^2 \, (E_{max} - E_{min})
\label{Eq:I_analytic}
\end{equation}
where we recall that $E_{min}$$=$max\{$E_{CT},E_{FT}$\},
$E_{max}$$=$min\{$E_{VB},E_{FB}$\} define the tunneling window.

It should be noticed that Eq.\ref{Eq:I_analytic} is consistent with
a complete loss of momentum conservation, so that the current is
simply proportional to the integral over the tunneling window of the
product of the density of states in the two 2D layers. Since for a
parabolic effective mass approximation the density of states is
energy independent, the current turns out to be simply proportional
to the width of the tunneling window.
In physical terms, Eq.\ref{Eq:I_analytic} corresponds to a situation
where the scattering produces a complete momentum randomization 
during the tunneling process.

As can be seen, as long as the top layer is {\it not} degenerate we
have $E_{min}$$=$$E_{CT}$ and the tunneling window widens with the
increase of the top gate voltage V$_{T,G}$, hence according to
Eq.\ref{Eq:I_analytic} the current is expected to increase linearly
with V$_{T,G}$.
However, when the tunneling window increases to such an extent that
$q^2_{M}$ becomes comparable to or larger than 1$/$$L_c^2$, then
part of the $q$ values in the integration of Eq.\ref{Eq:Analytic3}
belong to the tail of the spectrum S$_{F}$(q) defined in
Eq.\ref{Eq:Scr}, and so their contribution to the current becomes
progressively vanishing. The corresponding physical picture is that, 
while the tunneling window increases, the magnitude of the \wavevs\
in the two 2D layers also increases, and consequently the scattering
can no longer provide momentum randomization for all the possible \wavevs\
involved in the tunneling process.
Under these circumstances the current is expected to first increase
sub-linearly with V$_{TG}$ and eventually saturate for large enough
V$_{TG}$ values.

\section{Numerical results for the tunneling current
}\label{Sec:Results}

The 2D materials used for the tunneling current calculations
reported in this paper are the hexagonal monolayer $MoS_2$ and
$WTe_2$. The band structure for $MoS_2$ and $WTe_2$ have been
calculated by using a density functional theory (DFT) approach
\cite{gong2013band,liu2013three}, showing that these materials have a direct
bandgap with the band edges for both the valence and
the conduction band residing at the $K$ point in the 2D Brillouin
zone.
Fig.\ref{Fig:ParabolicBand} shows that in a range of about 0.4 eV
from the band edges the DFT results can be fitted fairly well by
using an energy relation based on a simple parabolic effective mass
approximation (dashed lines). Hence the parabolic effective mass
approximation appears adequate for the purposes of this work, which
is focussed on a device concept for extremely small supply
voltages ($<$\,0.5 V). The values for the effective masses inferred from the
fitting of the DFT calculations are tabulated in
Tab.\ref{Tab:MaterialParameter}
together with some other material parameters relevant for the tunneling current calculations.

In all current calculations we assume a top gate work function of 4.17
eV (Aluminium) and back gate work function of 5.17 eV (p++ Silicon) and
the top and bottom oxide have an effective oxide thickness (EOT) of 1 nm (see
Fig.\ref{Fig:Device_BS}). The top 2D layer consists of hexagonal
monolayer MoS$_2$ while the bottom 2D layer is hexagonal monolayer
WTe$_2$.
An $n$-type and $p$-type doping density of $10^{12} cm^{-2}$ by impurities and full ionization are
assumed respectively in the top and bottom 2D layer and the relative
dielectric constant of the interlayer material is set to 4.2 (e.g.
boron nitride). The voltage  $V_{DS}$ between the drain and the
source is set to 0.3 V and the back gate is grounded for all
calculations, unless otherwise stated.

In Fig. \ref{Fig:BA&J_Lc-L}, the results of numerical calculations are shown for the band alignment and the current density versus
the top gate voltage $V_{TG}$. Figure \ref{Fig:BA&J_Lc-L}(a) shows that
the top gate voltage can effectively govern the band alignment in
the device and, in particular, the crossing and uncrossing between
the conduction band minimum E$_{CT}$ in the top layer and the
valence band maximum E$_{VB}$ in the bottom layer, which
discriminates between the on and off state of the transistor.

\begin{figure}[!ht]
\centering
\subfigure[]{
\includegraphics[width=0.48\textwidth]{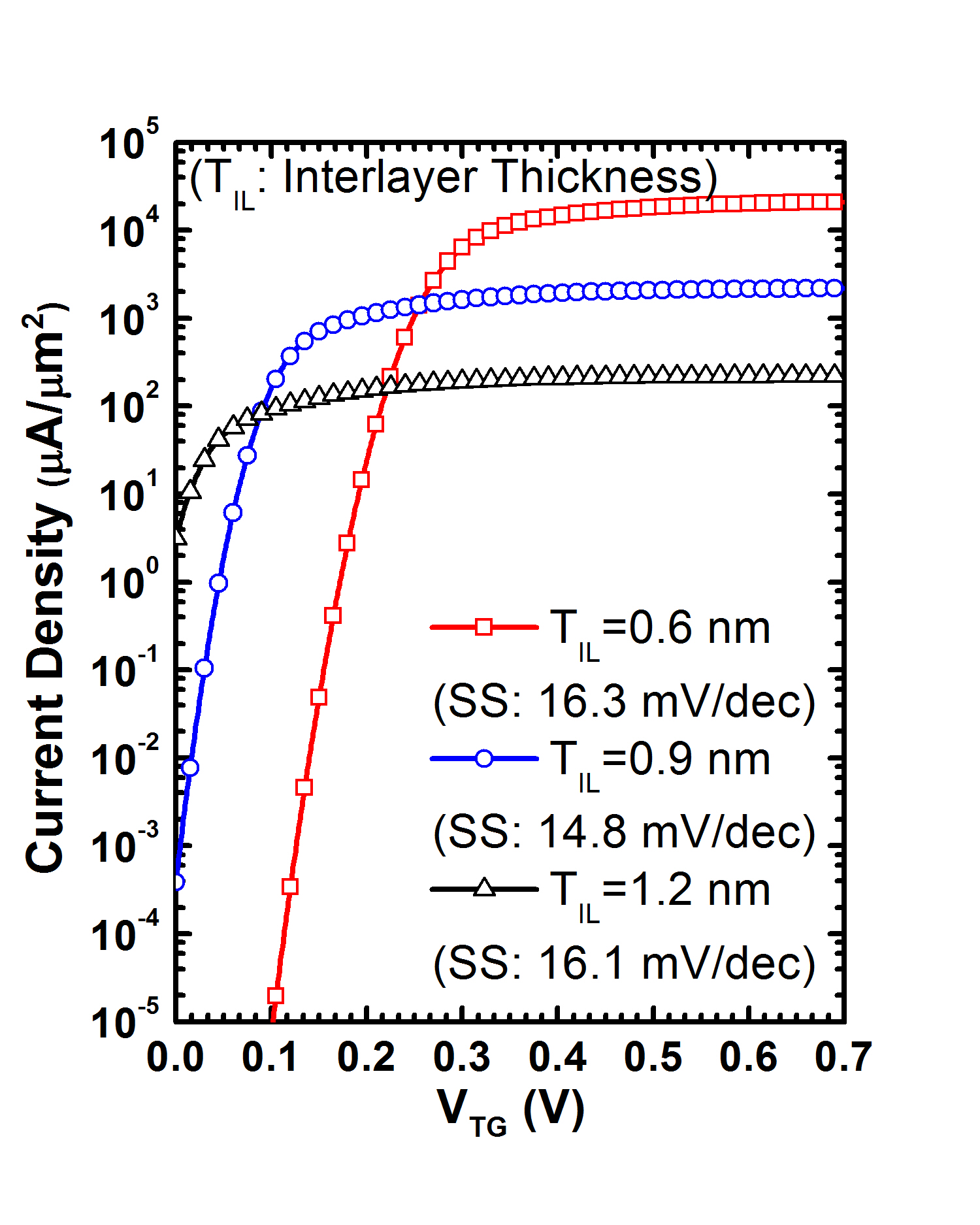}
}
\subfigure[]{
\includegraphics[width=0.48\textwidth]{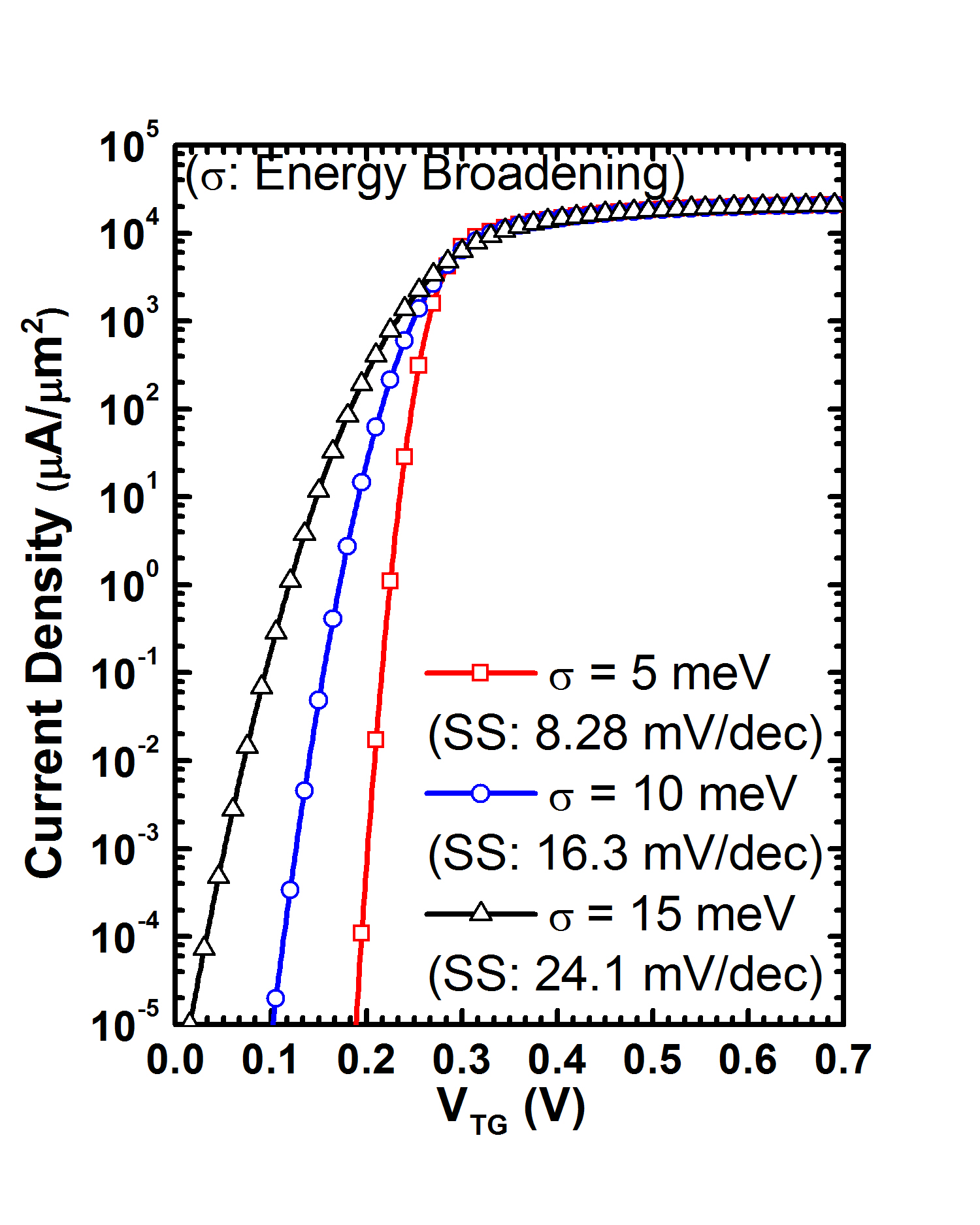}
} \caption[x]{ Numerical calculations for: (a) current density versus $V_{TG}$ with several
interlayer thicknesses; (b) current density versus $V_{TG}$ with different values of energy broadening $\sigma$. The matrix element is $M_{B0}=0.1\, eV$;
the decay constant of \wavef\ in the interlayer is $\kappa=3.8\, nm^{-1}$. In (a)
the energy broadening is $\sigma=10\, meV$. In (b) the interlayer
thickness is $T_{IL}=0.6 \, nm$ (e.g. 2 atomic layers of BN). $V_{BG}=0$ and $V_{TG}=0.3\,V$ in both (a) and (b).}
\label{Fig:Steepness}
\end{figure}

The I$_{DS}$ versus V$_{TG}$ characteristic in
Fig.\ref{Fig:BA&J_Lc-L}(b) can be roughly divided into three
different regions: sub-threshold region, linear region and
saturation region. The sub-threshold region corresponds to the
condition E$_{CT}$$>$E$_{VB}$ (see also Fig.\ref{Fig:BA&J_Lc-L}(a)),
where the very steep current dependence on V$_{TG}$ is illustrated better in Fig.\ref{Fig:Steepness} and  will be discussed below.

In the second region I$_{DS}$ exhibits an approximately linear
dependence on V$_{TG}$, in fact the current is roughly proportional
to the energy tunneling window, as discussed in
Sec.\ref{Sec:AnalyticResults} and predicted by
Eq.\ref{Eq:I_analytic}, because the tunneling window is small enough
that the condition $q^2_{M}$$\ll$1$/$$L_c^2$ is fulfilled. In this
region I$_{DS}$ is proportional to the long-wavelength part of
scattering spectrum (i.e. small $q$ values), hence the current
increases with L$_c$, as expected from Eq.\ref{Eq:I_analytic}. The
super-linear behavior of I$_{DS}$ at small V$_{TG}$ values observed
in Fig.\ref{Fig:BA&J_Lc-L}(b) is due to the tail of the Fermi
occupation function in the top layer. 
When V$_{TG}$ is increased above approximately 0.5V, the current in
Fig.\ref{Fig:BA&J_Lc-L}(b) enters the saturation region, where I$_{DS}$ increasing with V$_{TG}$ slows down because of the decay of the scattering spectrum S$_R(q)$ for $q$
values larger than $1$$/$$L_c$ (see Eq.\ref{Eq:Scr}). 

In Fig.\ref{Fig:Steepness} we analyze the I-V curves for different
interlayer thicknesses T$_{IL}$ and broadening energies $\sigma$;
in all cases an average inverse sub-threshold slope (SS)
is extracted in the I$_{DS}$ range from 10$^{-3}$ and
1 [$\mu$A$/$$\mu m^2$]. Figure \ref{Fig:Steepness}(a) shows
that the tunneling current increases exponentially by decreasing
T$_{IL}$, and the decay constant $\kappa$$=$3.8 nm$^{-1}$ employed in
our calculations results in a dependence on T$_{IL}$ that is roughly
consistent with the dependence experimentally reported in graphene
based interlayer tunneling devices \cite{britnell2012atomically}.
The threshold voltages are also shifted to lower values by increasing T$_{IL}$.
It can be seen that the T$_{IL}$ impact on SS
is overall quite modest and for all the T$_{IL}$ values the
simulations indicate a very steep I-V curve in the sub-threshold region ($<$\,20 mV/dec). 

Figure \ref{Fig:Steepness}(b) shows that according to the model
employed in our calculations SS is mainly governed by the parameter
$\sigma$ of the energy broadening (Eq.\ref{Eq:Distri.}). This result is expected, as already mentioned in Sec.\ref{Sec:broadening}, since in our model the
energy broadening is the physical factor setting the minimum value for SS  and the I$_{DS}$ versus V$_{TG}$ approaches a step-like curve when $\sigma$ is zero due to the step-like DoS of these 2D semiconductors \cite{agarwal2011pronounced}.
These results suggest that the energy
broadening in the 2D materials plays a very critical role in
achieving experimentally low SS values in the proposed Thin-TFETs.

\section{Discussion and conclusions}\label{Sec:DiscussConclusion}

This paper proposed a new steep slope transistor based on the
interlayer tunneling between two 2D semiconductor
materials and presented a detailed model to
discuss the physical mechanisms governing the device operation and
to gain an insight about the tradeoffs implied in the design of the
transistor.

The tunnel transistor based on 2D semiconductors has the potential for a
very steep subthreshold region and the subthreshold swing is
ultimately limited by the energy broadening in the two 2D materials.
The energy broadening can have different physical origins such as
disorder, charged impurities in the 2D layers or in the surrounding
materials \cite{ghazali1985disorder} $^,$ \cite{sarma1982thomas},
phonon scattering \cite{bockelmann1990phonon} and microscopic
roughness at interfaces \cite{knabchen1995self}.
In our calculations we accounted for the energy broadening by
assuming a simple gaussian energy spectrum with no explicit
reference to a specific physical mechanism. However, a more detailed
and quantitative description of the energy broadening is instrumental in physical modeling of the device and its design. 

Quite interestingly, our analysis suggests that, while a possible
rotational misalignment between the two 2D layers can affect the
absolute value of the tunneling current, the misalignment is not expected to
significantly degrade the steep subthreshold slope, which is the
crucial figure of merit for a steep slope transistor.

\begin{figure}[!ht]
\centering
\subfigure[]{
\includegraphics[width=0.48\textwidth]{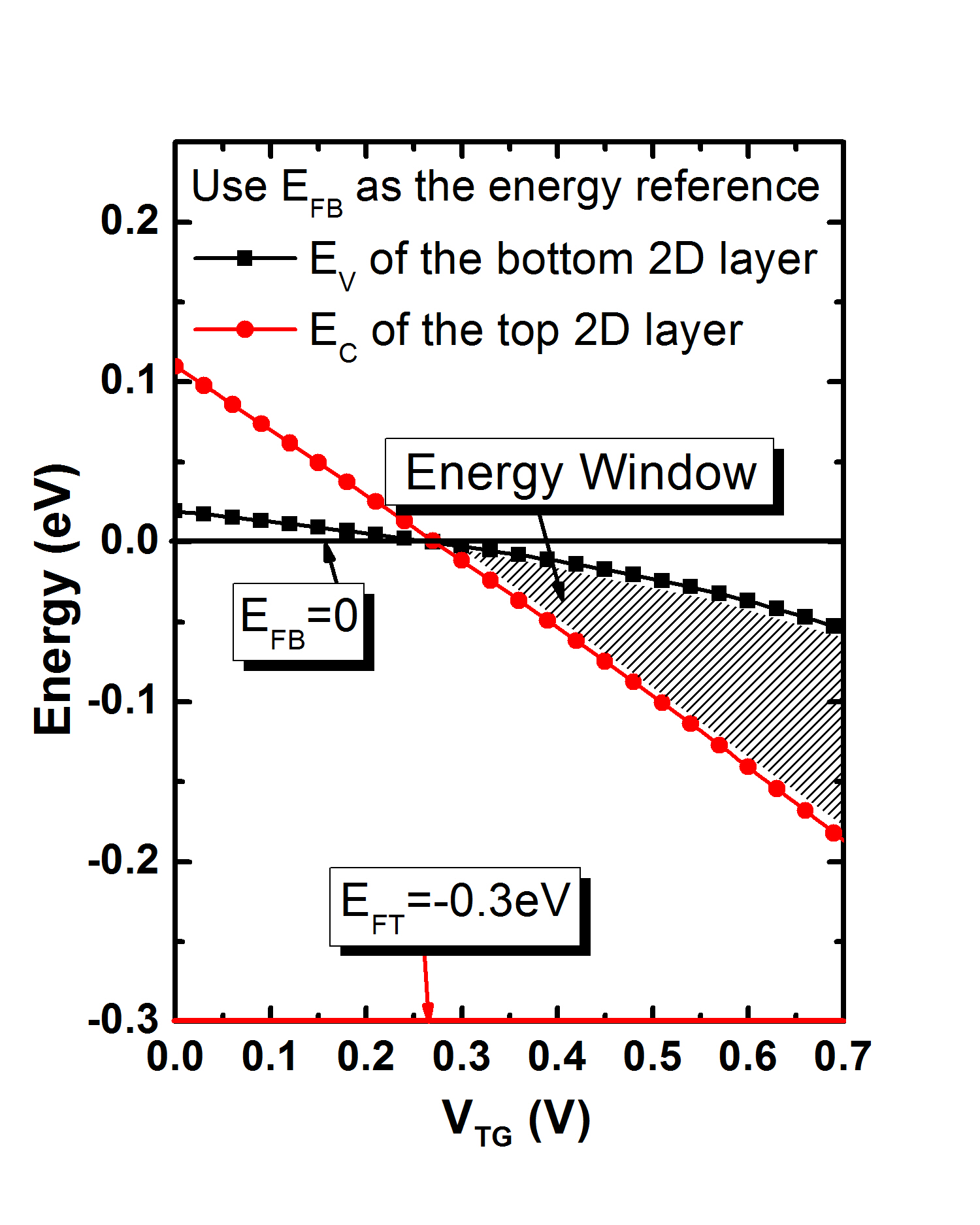}
}
\subfigure[]{
\includegraphics[width=0.48\textwidth]{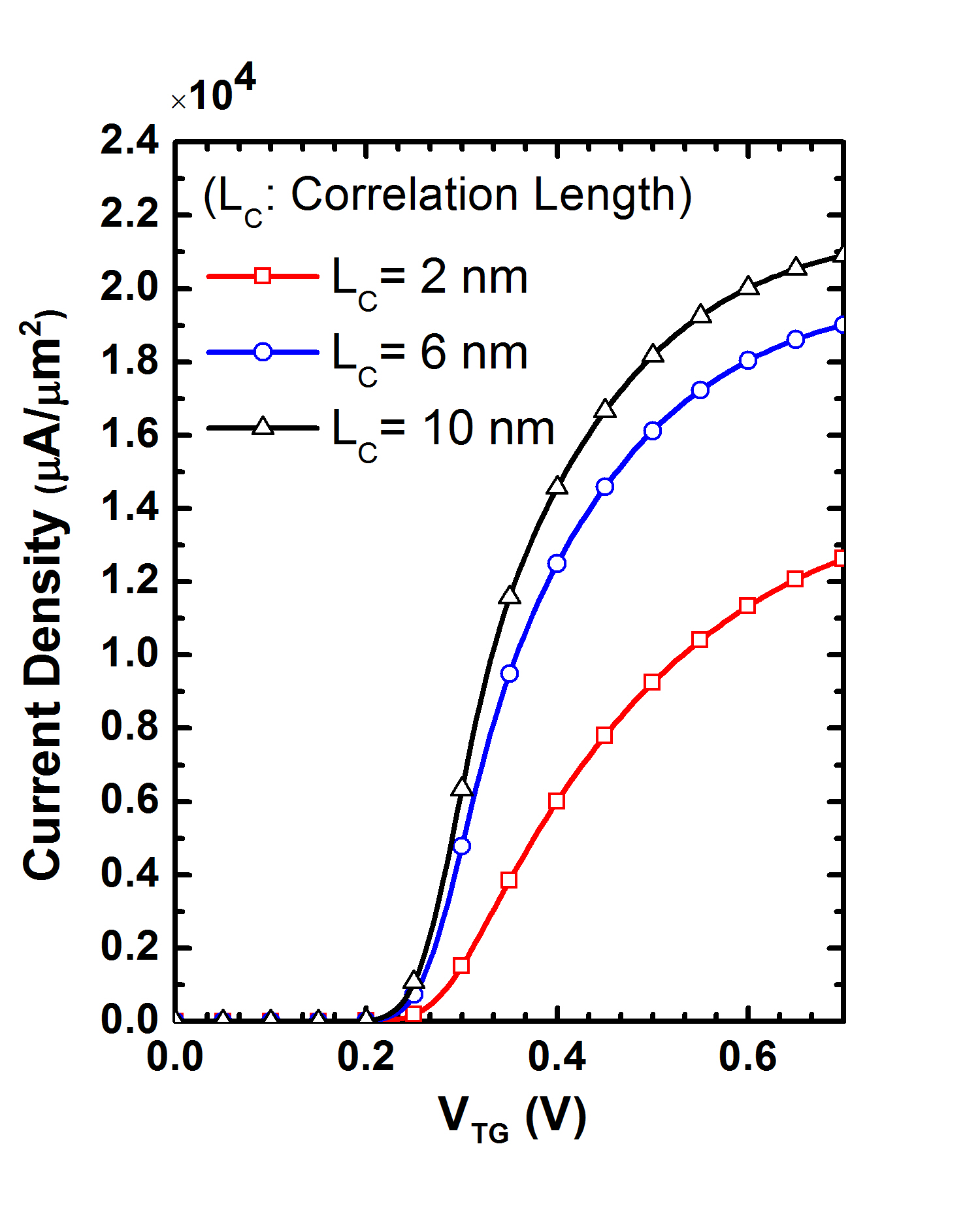}
} \caption[x]{ Numerical results of (a) band alignment versus the top gate voltage $V_{TG}$ and (b) tunnel current
density versus the top gate voltage $V_{TG}$ for different values of the correlation length L$_C$. The parameters used in (b) are: matrix element is $M_{B0}=0.1\, eV$; decay constant of \wavef\ in the interlayer is $\kappa=3.8\, nm^{-1}$; energy
broadening is $\sigma=10\, meV$ and interlayer thickness is $T_{IL}=0.6 \,
nm$ (e.g. 2 atomic layers of BN). $V_{BG}=0$ and $V_{DS}=0.3\,V$ in both (a) and (b).}
\label{Fig:BA&J_Lc-L}
\end{figure}

An optimal operation of the device demands a good electrostatic control of
the top gate voltage V$_{TG}$ on the band alignments in the material
stack,
as shown for example in Fig.\ref{Fig:BA&J_Lc-L}(a), which may become problematic if the electric filed in the interlayer is effectively screened by the high electron concentration in the top 2D layer. 
Consequently, since high carrier concentrations in the 2D layers are essential to reduce the layer resistivities, a tradeoff exists between the gate control and layer resistivities; as a result, doping concentrations in these 2D layers are important design parameters in addition to tuning the threshold voltage.
In this respect, chemical doping of TMD materials have been recently
demonstrated \cite{fang2013degenerate,fang2012high},
however these doping technologies are still far less mature than
they are for 3D semiconductors, and improvements in in-situ doping
will be very important for optimization of the device
performance.
Since our model does not include the lateral transport in the 2D
materials, an exploration of the above design tradeoffs goes beyond
the scope of the present paper and demands the development of more
complete transport models.

The transport model proposed in this work does not account for
possible traps or defects assisted tunneling, which have been
recently recognized as a serious hindrance to the experimental
realization of Tunnel-FETs exhibiting a sub-threshold swing better
than 60 mV/dec \cite{pala2013interface,esseni2013interface}. A large density of
states in the gap of the 2D materials may even lead to a Fermi
level pinning that would drastically degrade the gate control on the
band alignment and undermine the overall device operation. In this
respect, from a fundamental viewpoint the 2D crystals may offer
advantages over their 3D counterparts because they are
inherently free of broken/dangling bonds at the interfaces
\cite{jena2013tunneling}. However, the fabrication technologies for 2D
crystals are still in an embryonal stage compared to technologies
for conventional semiconductors, hence the control of defects in the
2D materials will be a challenge for the development of the
proposed tunneling transistor.

The simulation results reported in this paper indicate that the
newly proposed transistor based on interlayer tunneling between two
2D materials has the potential for a very steep turn-on
characteristic, because the vertical stack of 2D materials
having an energy gap is probably the device structure that allows
for the most effective, gate controlled crossing and uncrossing
between the edges of the bands involved in the tunneling process.
Our modeling approach based on the Bardeen's transfer Hamiltonian
is by no means a complete device model but
instead a starting point to gain insight about its working
principle and its design.
At the present time an experimental
demonstration of the device appears of crucial importance, first of all to validate the device concept,
and then to help estimate the numerical value of a few parameters in the transport model that can be determined only by comparing to experiments. 

\begin{figure}[!ht]
\centering \subfigure[]{
\includegraphics[width=0.48\textwidth]{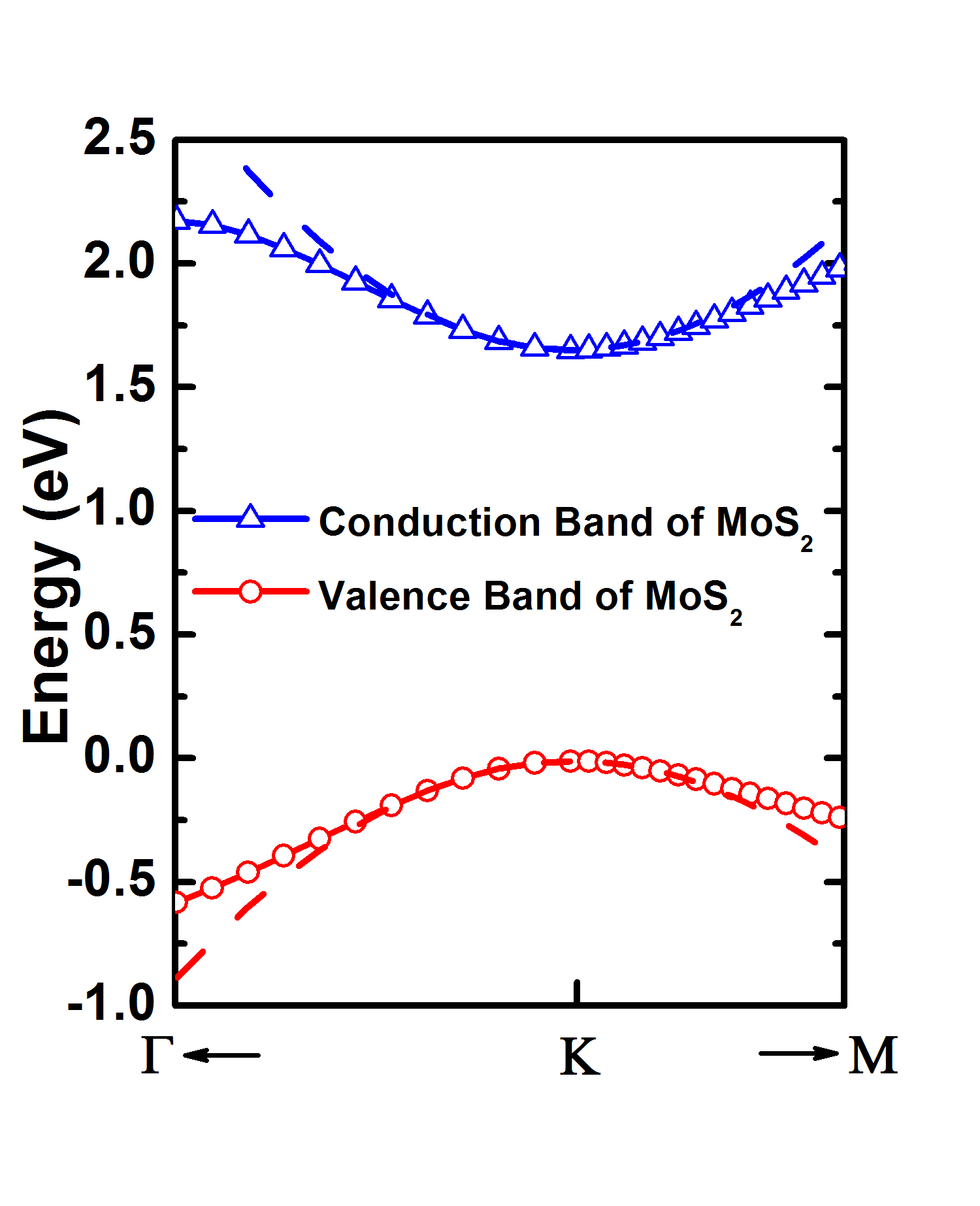}
} \subfigure[]{
\includegraphics[width=0.48\textwidth]{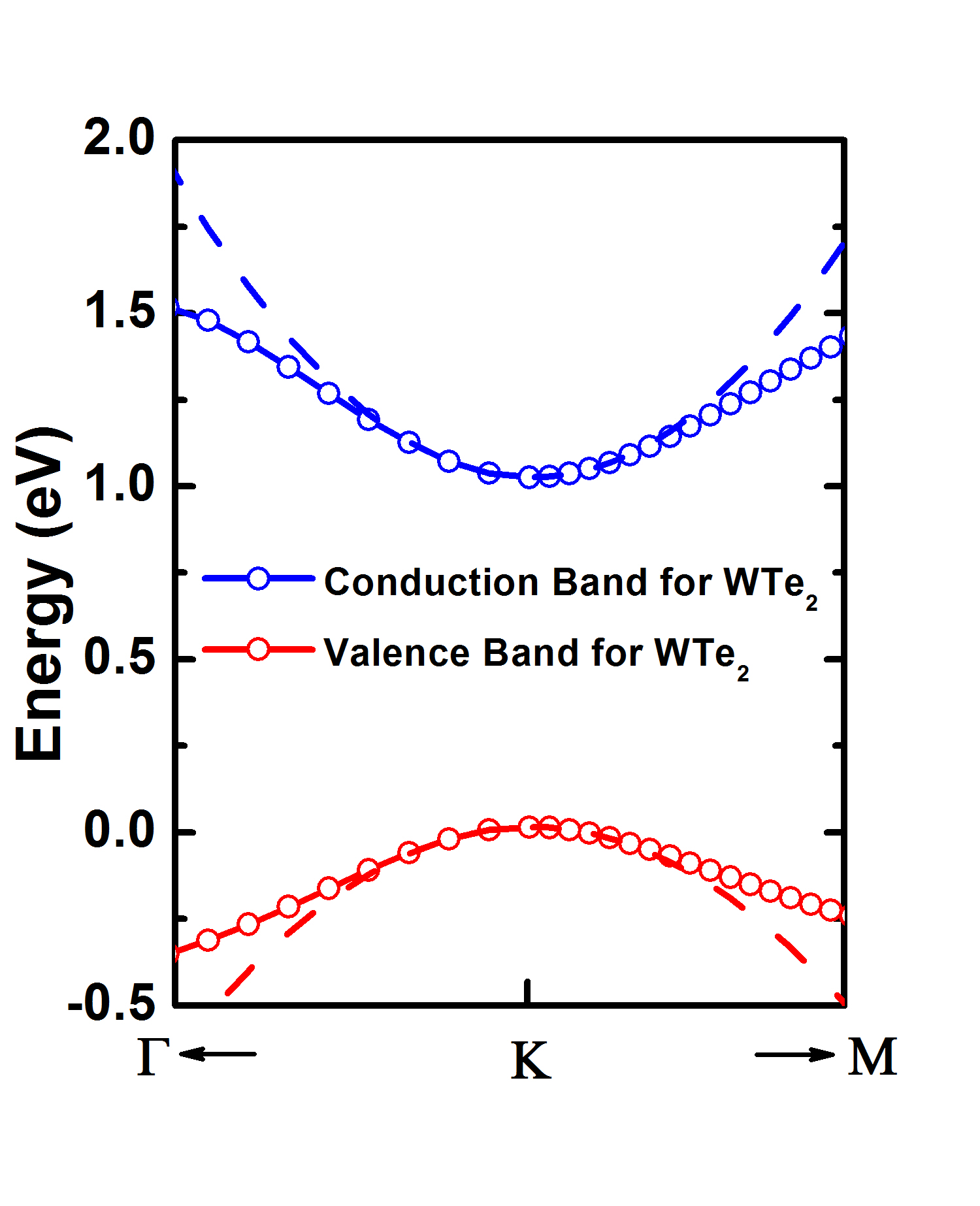}
} \caption[x]{ 
(a) Band structure for hexagonal monolayer MoS$_2$ and (b) hexagonal monolayer WTe$_2$ as obtained using DFT method described in the paper of C. Gong et.al. \cite{gong2013band}.
The dashed lines represent the analytical approximation obtained
with a parabolic effective mass model.}
\label{Fig:ParabolicBand}
\end{figure}

{\bf Acknowledgments:} This work was supported in part by the Center
for Low Energy Systems Technology (LEAST), one of six SRC STARnet
Centers, sponsored by MARCO and DARPA, by the Air Force Office of Scientific Research (FA9550-12-1-0257), and by a Fulbright Fellowship for D. Esseni. The authors are also grateful for the helpful discussions with Profs. K. J. Cho, R. Feenstra and A. Seabaugh. The authors are especially thankful to Dr. Cheng Gong in Dr. K. J. Cho's group for providing the calculated band structure data shown in Fig. \ref{Fig:ParabolicBand}.

\bibliography{JAP2013}

\newpage
\begin{table}[h]
\begin{center}
\begin{tabular}{|c|c|c|c|c|c|}
\hline
  & {\bf Bandgap (eV)}  & {\bf  Electron affinity ($\chi$)} & {\bf Conduction band}            & {\bf Valence band}  \\
  &             &                           & {\bf  effective mass (m$_c$)} & {\bf effective mass (m$_v$)} \\
\hline
{\bf MoS$_2$} &  1.8  &  4.30  &  0.378   &  0.461   \\\hline
{\bf WTe$_2$} &  0.9  &  3.65  &  0.235   &  0.319   \\\hline
\end{tabular}
\vspace{4mm} \caption{The band gaps, electron affinities and effective masses used for MoS$_2$ and WTe$_2$ }
\label{Tab:MaterialParameter}
\end{center}
\end{table}

\newpage

\newpage

\newpage

\newpage

\newpage

\newpage

\end{document}